\documentclass[12pt]{article}
\usepackage{graphicx}
\usepackage{cite}
\newcommand{\vev}[1]{ \left\langle {#1} \right\rangle }

\newcommand{\beq}{\begin{equation}}
\newcommand{\eeq}{\end{equation}}

\begin{document}

\begin{titlepage}

\begin{center}

%\vspace{2cm}
{\hbox to\hsize {\hfill UFIFT-HEP-04-5 }}
{\hbox to\hsize {\hfill CLNS04/1872 }}
{\hbox to\hsize {\hfill UCB-PTH-04/11 }}
{\hbox to\hsize {\hfill  LBNL-54921 }}
\medskip

%\vspace{2cm}

{\Large \bf Little Supersymmetry and the \\
\medskip
Supersymmetric Little Hierarchy Problem
%\footnote{This
%work was supported in part by the
%Director, Office of Science, Office of High Energy and Nuclear
%Physics, Division of High Energy Physics of the U.S. Department of
%Energy under Contract DE-AC03-76SF00098, in part by the National
%Science Foundation under grants PHY-0098840.}
}

\bigskip

{\bf Andreas Birkedal}$^{\bf a,b,e}$,
{\bf Z. Chacko}$^{\bf c,d,f}$, 
and {\bf Mary K. Gaillard}$^{\bf c,d,g}$ \\

\bigskip
$^{\bf a}${\small \it Physics Department, University of Florida,
Gainesville, FL 32611, USA \\
\smallskip
$^{\bf b}$ Institute for High-Energy Phenomenology, Cornell University, \\
Ithaca, NY 14853, USA \\
\medskip
$^{\bf c}$ Department of Physics, University of California, Berkeley, CA 94720, USA 
\\
\smallskip
$^{\bf d}$ Theoretical Physics Group, Lawrence Berkeley National Laboratory,\\
Berkeley, CA 94720, USA \\
\medskip
$^{\bf e}$andreasb@mail.lns.cornell.edu, $^{\bf 
f}$zchacko@thsrv.lbl.gov, $^{\bf g}$mkgaillard@lbl.gov}
\bigskip

{\bf Abstract}

\end{center}
\noindent

%{\tiny 
The current experimental lower bound on the Higgs mass significantly restricts the
allowed parameter space in most realistic supersymmetric models, with the
consequence that these models exhibit significant fine-tuning. We propose a
solution to this `supersymmetric little hierarchy problem'. We consider scenarios
where the stop masses are relatively heavy - in the 500 GeV to a TeV range.
Radiative stability of the Higgs soft mass against quantum corrections from the top
quark Yukawa coupling is achieved by imposing a global SU(3) symmetry on this
interaction. This global symmetry is only approximate - it is not respected by the
gauge interactions. A subgroup of the global symmetry is gauged by the familiar
SU(2) of the Standard Model. The physical Higgs is significantly lighter than the
other scalars because it is the pseudo-Goldstone boson associated with the breaking
of this symmetry. Radiative corrections to the Higgs potential naturally lead to
the right pattern of gauge and global symmetry breaking.  We show that both the
gauge and global symmetries can be embedded into a single SU(6) grand unifying
group, thereby maintaining the prediction of gauge coupling unification. Among the
firm predictions of this class of models are new states with the quantum numbers of
10 and $\overline{10}$ under SU(5) close to the TeV scale. The Higgs mass is
expected to be below 130 GeV, just as in the MSSM.
%}

\end{titlepage}

\renewcommand{\thepage}{\arabic{page}}
\setcounter{page}{1}

\section{Introduction}

Supersymmetry is perhaps the most attractive solution of the hierarchy problem. Among
its many appealing features is the observed unification of the Standard Model gauge
couplings {\cite{DG}} normalized as in
SU(5){\cite{unification}} in the Minimal Supersymmetric Standard Model or `MSSM'.
However, most realistic supersymmetric models today suffer from a naturalness problem,
called the `supersymmetric little hierarchy problem'. This problem arises because in
most of the parameter space of these theories the Higgs mass lies below the current
experimental lower bound. As a consequence of this these theories are typically
fine-tuned at the level of a few percent, rendering them highly unnatural.

Let us attempt to understand in more detail the origin of the supersymmetric little
hierarchy problem. At tree level in the MSSM, the mass of the lightest neutral Higgs
is bounded by M$_Z$. Therefore if the lightest neutral Higgs is to be heavy enough to
avoid the current experimental bound, there must be significant quantum corrections to
the Higgs potential, and in particular to the Higgs quartic coupling. Loops involving
the scalar superpartners of the top quark, the stops, do indeed generate such a
correction but only if the stops are relatively heavy. However, loops involving the
stops also give quantum corrections to the Higgs soft mass parameter, and if the stops
are heavy these corrections are too large and must be cancelled to within a few
percent against the $\mu$ term to obtain correct electroweak symmetry breaking. This
leads to significant fine-tuning. Therefore we see that the source of the problem is
that the stop masses are being required to meet two contradictory criteria
\begin{itemize} 
\item 
they must be large to generate a sizable quartic correction to
the Higgs potential 
\item 
they must be small to avoid generating sizable logarithmic
corrections to the Higgs soft mass parameter. 
\end{itemize} 
The fine-tuning is the result of an attempt to balance these two effects. Although
both contributions arise from loops involving the top Yukawa couplings, the correction
to the quartic arises mainly from loop momenta below the stop mass, while the unwanted
correction to the soft mass parameter arises mainly from loop momenta much larger than
the scale of the soft masses.  Therefore if the Higgs mass can be protected against
radiative corrections from scales significantly higher than the superpartner masses,
the supersymmetric little hierarchy problem can be solved.

More insight into the problem may be gained by considering the manner in which
supersymmetry addresses the hierarchy problem.  We start with the observation that at
the one loop level there are in fact three separate hierarchy problems in the Standard
Model -
\begin{itemize}
\item
quadratic corrections to the Higgs mass squared associated with the Standard Model 
gauge
couplings
\item
quadratic corrections to the Higgs mass squared associated with the Higgs self 
coupling
\item
quadratic corrections to the Higgs mass squared associated with its Yukawa 
couplings to 
the
Standard Model fermions.
\end{itemize}
In supersymmetric models the first two of these are closely related because the Higgs
quartic term arises from the D-terms of the gauge multiplets. However they are indeed
distinct from the third. This can be seen from the fact that quadratic divergences
arising from the Standard Model Yukawa couplings are cancelled by the scalar
superpartners of the fermions while divergences arising from the Higgs gauge and self
interactions are cancelled by the gauginos and Higgsinos. The source of the 
supersymmetric little hierarchy problem is that the bound on the Higgs mass implies
that the stop masses must be heavy. This then creates a fourth, much milder hierarchy
problem.
\begin{itemize}
\item  
logarithmic corrections to the Higgs mass squared associated with the Yukawa 
couplings to 
the top quark, and proportional to the stop mass squared
\end{itemize}
This fourth hierarchy problem is clearly more closely related to the last in the
list of three above than to the first two. For example, by keeping the gauginos
light it is possible to address the first pair of hierarchy problems without
addressing the second pair. 

In this paper we propose a new solution to the supersymmetric little hierarchy
problem. We consider a scenario where the stop masses are relatively heavy, and lie
in the 500 GeV to 1 TeV range. In order to achieve radiative stability against the
quantum corrections from the top Yukawa coupling a global symmetry, which for
concreteness we will take to be SU(3$)_{\rm L}$ is imposed on this interaction. The
global symmetry is only approximate - it is not respected by the gauge
interactions.  A subgroup of the global symmetry is gauged by the SU(2$)_{\rm L}$
in the Standard Model. The Higgs fields of the MSSM emerge from multiplets which
transform as a {\bf $3$} and {\bf $\bar{3}$} under the global symmetry. One linear
combination of the two SU(2)${\rm _L}$ doublets in this {\bf $3$}, {\bf $\bar{3}$}
pair, which is the physical Higgs, is protected against quantum corrections from
the Yukawa couplings because it is the pseudo-Goldstone boson associated with the
breaking of this global symmetry. This model is in the spirit of `little Higgs'
theories ({\cite{littleHiggs}, {\cite{KS}}, see also {\cite{Georgi}}), which solve
the little hierarchy problem by generating a light Standard Model Higgs as the
pseudo-Goldstone boson of an approximate global symmetry\footnote{ Earlier work on
using global symmetries to protect against radiatively induced scalar masses in a
locally supersymmetric context can be found in {\cite{GHZAPR}}.}. A clear and
concise introduction to these theories may be found in {\cite{Schmaltz}}.

If at some higher scale where the supersymmetry breaking masses are generated the
global symmetry is exact, the scalar masses at this scale will respect SU(3). When
renormalization group evolved down to the scale of the scalar masses they will
still respect the symmetry to a very good approximation because at one loop nearly
all logarithmically enhanced corrections to the SU(3) invariant form are
proportional to the square of the electroweak gaugino masses, which are typically
small. At one loop the only SU(3) violating correction that does not fall into
this category is proportional to the sum of all the scalar masses squared weighted
by hypercharge. However, this quantity vanishes or is small in a large class of
models, either as a consequence of the pattern of supersymmetry breaking, or as a
consequence of the way in which the model is embedded in a unified theory. Since
the form of the Higgs potential at the scale of the scalar masses is approximately
SU(3) symmetric, the Standard Model Higgs, which is the pseudo-Goldstone
associated with the breaking of this global symmetry, will be significantly
lighter than the other scalars and can be made to naturally acquire a weak scale
VEV.

How is the SU(3) global symmetry broken? The simplest possibilty is that it is
broken radiatively in exactly the same manner as conventional electroweak symmetry
breaking in the MSSM driven by the top Yukawa coupling. A quartic restoring
potential for the Higgs of the right form may be obtained by incorporating into the
model a U(1) gauge symmetry that commutes with the global SU(3). The right alignment
pattern is favored because of the D terms associated with the Standard Model gauge
interactions that violate the global symmetry. The subsequent breaking of the
electroweak symmetry group down to U(1) electromagnetism also occurs naturally.
Quantum corrections from stop loops to the Higgs quartic potential now allow the
mass of the lightest neutral Higgs to be above the experimental bound without
significant fine tuning.

Is the prediction of gauge coupling unification maintained in this class of models?
We show that it is straightforward to embed the global SU(3) in an SU(6) grand
unifying group. The Standard Model quarks, leptons and gauge fields then have their
conventional embedding in the SU(5) subgroup of SU(6). In order to realize the SU(3)
symmetry at low scales while preserving the prediction of gauge coupling
unification, a pair of fields with the quantum numbers of a 10 and $\overline{10}$
under SU(5) must be present at or close to the weak scale. This is a firm prediction
of this model.

What are the characteristic features of such a `little supersymmetric' theory ?  
We expect the stops to be relatively heavy, above about 500 GeV. The constraints
on the rest of the superpartner spectrum are fewer, but natural electroweak
symmetry breaking requires that the electroweak gauginos be lighter than the
stops.  There is an additional Z' gauge boson associated with the new U(1) gauge
symmetry that couples to the Standard Model fields, and which typically has a mass
in the 300 GeV to a TeV range. There are also new states with masses of order a
TeV associated with the global SU(3) symmetry of the top Yukawa. These have
vector-like charges under the Standard Model gauge symmetry, and come in complete
SU(5) multiplets in order to preserve unification.  In contrast to other proposed
solutions of the supersymmetric little hierarchy problem, the mass of the lightest
neutral Higgs is expected to be below 130 GeV, just as in the MSSM.

While recently there has been considerable attention focused on potential
solutions of the supersymmetric little hierarchy problem {\cite{Su}},
{\cite{Kobayashi}}, the approach we follow here differs significantly from
that of other authors.  Most recent work relies on altering the form of the Higgs
potential close to the weak scale with a view to invalidating the MSSM bound on
the Higgs mass. For example, by generating additional quartic terms beyond those
in the MSSM the Higgs mass can be raised above 130 GeV ameliorating the fine
tuning problem. For early work in this direction see, for example {\cite{Haber}}.  
In contrast, we attempt to make the mass of the physical Higgs insensitive to loop
corrections from higher scales. For an alternative recent approach where
superconformal symmetry is used to suppress the Higgs mass parameters relative to
the other soft parameters, see {\cite{Kobayashi}}.

\section{A Minimal Model}

\subsection{The Global Symmetry}

The Yukawa coupling of the top quark to the up-type Higgs in the MSSM has the
familiar form  
\begin{equation}
(3,2)_{Q_3} (1,2)_{H_u} (\bar{3},1)_{{t^c}}  
\end{equation}
where the numbers in brackets indicate the quantum numbers of the various fields
under SU(3$) \times$ SU(2$)_{\rm L}$. For simplicity, we have suppressed the
hypercharge quantum numbers. Here, following the usual convention, $Q_3$ represents
the third generation left handed quarks, ${t^c}$ represents the third generation
left handed antiquark and $H_u$ is the up-type Higgs. We wish to suitably extend
this interaction to make it invariant under a global SU(3) symmetry, 
which we denote by G, of which the familiar
SU(2$)_{\rm L}$ is a gauged subgroup. This can be done by extending
\begin{equation}
\label{top}
(3,2)_{Q_3} (1,2)_{H_u} (\bar{3},1)_{{t^c}} \rightarrow 
(3,\bar{3})_{\hat{Q}_3} (1,3)_{\hat{H}_u} (\bar{3},1)_{{t^c}}
\end{equation}
where the second number in each bracket on the right hand side of the arrow
now indicates the transformation properties of each field under the new 
global SU(3) symmetry G. Here
\begin{eqnarray}
\hat{Q}_3 &=& \left( {Q}_3, \overline{T^c} \right) \\
\hat{H}_u &=& \left( {H}_u, S_u \right)
\end{eqnarray}     
where $S_u$ and $\overline{T^c}$ are the new states we have added to the theory
to make it invariant under G. $S_u$ 
is a singlet under the Standard Model gauge group while
the quantum numbers of $\overline{T^c}$ are such that it is vector-like with 
respect to $t^c$. 
We also embed the down-type Higgs $H_d$ into a $\bar{3}$ representation of the SU(3) 
global symmetry 
\begin{equation}
\hat{H}_d = \left( {H}_d, S_d \right)
\end{equation}
where $S_d$ is a singlet under the Standard Model gauge group. We further demand
that the soft masses for $\hat{H}_u$ and $\hat{H}_d$ as well as the $\mu$ and $B
\mu$ terms have a form which respects the global symmetry G at some high scale where
they are generated. However the SU(2$)_{\rm L} \times$ U(1$)_{\rm Y}$ gauge
interactions do not respect the global symmetry, and further we do not require that
the Yukawa couplings of the bottom quark or the lighter fermions
respect the global symmetry either, 
since they are relatively small. Then, since the SU(3) violating loop corrections 
to the
Higgs potential from the gauge and Yukawa couplings are relatively small, if
$S_u$ and $S_d$ acquire VEVs that break the approximate global symmetry one linear
combination of $H_u$ and ${H_d}^*$ will remain light as the pseudo-Goldstone boson
associated with the breaking of the approximate symmetry. This is the physical Higgs
field. We define $f$ as the scale at which the global symmetry is broken.
\begin{equation}
f = \sqrt{\vev{S_u}^2 + \vev{S_d}^2}
\end{equation}
If the global symmetry G is broken radiatively as in the MSSM we expect that the
scale $f$ is not far from the scale of the stop masses, which we denote by
$\hat{m}$. In particular, this will be the case if the coefficient of the quartic
restoring term in the Higgs potential for $S_u$ and $S_d$ is of order one. In our 
models this
restoring term arises from the D-term of a new U(1) gauge symmetry, which we
denote by U(1$)_{\rm E}$. The charges under U(1$)_{\rm E}$ are such that it commutes
with the global symmetry G, which then implies that the Standard Model fields are
charged under it. We can estimate $f$ more precisely as being of order
$\hat{m}/\hat{g}$, where $\hat{g}$ is gauge coupling strength of U(1$)_{\rm E}$.
Then radiative corrections to the Higgs soft masses from scales below $f$ that
violate the symmetry G are not large enough to significantly affect the
pseudo-Goldstone nature of the light Higgs.

There is a natural hierarchy between the scale $f$ at which the global symmetry is
broken and the scale at which electroweak symmetry is broken due to the difference
in the sizes of the terms in the Higgs potential which respect the global SU(3)
and those terms which violate the global SU(3). All the terms in the effective
theory for the pseudo-Goldstone field below the scale $\hat{m}$ arise from
relatively small loop effects that violate the global symmetry. It is therefore
these small effects which determine the manner in which electroweak symmetry is
broken. In the following subsections we explain in more detail how this happens.

Anomaly cancellation requires that there be an additional field $T^c$ with the
same gauge quantum numbers as $t^c$. In order to ensure that there are no unwanted
light states with masses below the weak scale we add to the theory a mass term of
the form $ \int d^2 \theta M T^c \overline{T^c} $. Although this term violates the
global symmetry G, provided that the mass $M$ is less than or of order $f$, the
global symmetry breaking scale, the SU(3) violating corrections to the Higgs
potential from scales below $f$ will not be large.  The physical left-handed third
generation antiquark of the Standard Model is a linear combination of $t^c$ and
$T^c$, with the exact ratio determined by the relative sizes of $f$ and $M$.

\subsection{The Pattern of Symmetry Breaking}

Under what circumstances does this model give the pattern of symmetry breaking we
seek, with $\vev{S_u}, \vev{S_d} \neq 0$, $\vev{{H}_u}, \vev{{H}_d} \neq 0$ and
$\vev{S_u} > \vev{{H}_u}$, $\vev{S_d} > \vev{H_d}$ ? In order to understand this we
consider the potential for the Higgs sector, $V_{\rm tot}$. Now $V_{\rm tot}$ can be
broken up into two parts, one part which respects the global SU(3) and which we
denote by $V_{\rm G}$, and another which breaks the global SU(3) and which we denote
by $V_{\rm B}$.
\begin{equation}
V_{\rm tot} = V_{\rm G} + V_{\rm B}
\end{equation}
The forms of $V_{\rm G}$ and $V_{\rm B}$ are:
\begin{eqnarray}
V_{\rm G} &=& 
% \left( \int d^2 \theta \; \hat{\mu} \hat{H}_u \hat{H}_d \right)
\left( \hat{\mu}^2 +{{\hat{m}}^2}_u \right) |{\hat{H}_u}|^2 
+ \left( \hat{\mu}^2 +{{\hat{m}}^2}_d \right) |{\hat{H}_d}|^2
+ \left( \hat{B} \; \hat{H}_u \; \hat{H}_d + {\rm h.c.} \right) 
\nonumber
\\
&&
+ \frac{\hat{g}^2}{8} \left( {|\hat{H}_u}|^2 - |{\hat{H}_d}|^2 \right)^2
\end{eqnarray}
\begin{eqnarray}
V_{\rm B} &=&
% \left( \int d^2 \theta \;{\mu} \tilde{H}_u \tilde{H}_d \right)
\left( 2 \delta{\mu} \; \hat{\mu} + \delta{\mu}^2 +{{\delta{m}}_u}^2 \right) 
|{{H}_u}|^2
+ \left( 2 \delta{\mu} \; \hat{\mu} + \delta{\mu}^2 +{{\delta{m}}_d}^2 \right) 
|{{H}_d}|^2
\nonumber
\\
&&
+ \left( \delta{B} {H}_u {H}_d + {\rm h.c.} \right)
+ \frac{{g_2}^2 + {g_1}^2}{8} \left( |{{H}_u}|^2 - |{{H}_d}|^2 \right)^2
+ \frac{{g_2}^2}{2} |H_u^* H_d|^2
\end{eqnarray}
where we have assumed for simplicity that all parameters in $V_{\rm tot}$
are real.
The terms involving the supersymmetric parameters $\mu$ and $\delta{\mu}$ above
emerge from the superpotential terms below which are generalizations of the $\mu$
term of the MSSM.
\begin{equation} 
W_{\rm tot} = \hat{\mu} \; \hat{H}_u \hat{H}_d + \delta{\mu} \; H_u H_d
\end{equation}
By assumption the dimensionful parameters in $V_{\rm G}$, which are expected to be 
of order $\hat{m}$,
are significantly larger
than those in $V_{\rm B}$. In the next sub-section it will be seen that 
this assumption is consistent with the sizes of the radiative 
corrections that violate the global SU(3), so that 
in most of parameter space 
$|{{\delta{m}}^2}| \ll |{{\hat{m}}^2}|$.

Stability of the Higgs potential $V_{\rm tot}$ for large field values requires
\begin{eqnarray}
{\label{stability}}
2\; \hat{\mu}^2 + \hat{m_u}^2 + \hat{m_d}^2 &>& 2 \; |B| \\
2\; \left( \hat{\mu} + \delta{\mu} \right)^2 + \hat{m_u}^2 + \delta{m_u}^2
+ \hat{m_d}^2 + \delta{m_d}^2 
&>& 2 \; |B + \; \delta B | .
\end{eqnarray}
We now argue that in regions of parameter space where both ${{\hat{m}}_u}^2 < 0$ and
${{\delta{m}}_u}^2 < 0$, but with $|{{\delta{m}}_u}^2| \ll |{{\hat{m}}_u}^2|$ we get
the pattern of symmetry breaking we require. We first consider minima of $V_{\rm
G}$, since the dimensionful parameters in $V_{\rm B}$ are significantly smaller. Now
the global SU(3) symmetry of $V_{\rm G}$ implies that it has both minima with
$\left[ \vev{S_u}, \vev{S_d} \neq 0 , \vev{H_u}, \vev{H_d} = 0 \right]$ as well as
minima with $\left[ \vev{S_u}, \vev{S_d} = 0 , \vev{H_u}, \vev{H_d} \neq 0 \right]$.
However the latter vacua are strongly disfavored by the D terms of SU(2) $\times$ 
U(1)
in $V_{\rm B}$. Therefore $\left[ \vev{S_u}, \vev{S_d} \neq 0 , \vev{H_u}, \vev{H_d}
= 0 \right]$ is an approximate solution to the potential $V_{\rm tot}$, if the
dimensionful terms in $V_{\rm B}$ are ignored. In order to determine the pattern of
electroweak symmetry breaking, we expand about this vacuum and determine the
effective theory for the light fields below the scale $\hat{m}$. This is an
expansion in the small parameter $(\delta{m}/\hat{m})^2$. The only field with mass
significantly smaller than $\hat{m}$ is one linear combination of $H_u$ and 
$H_d^*$,
which is the pseudo-Goldstone associated with the breaking of the SU(3) global
symmetry. We denote this light field by $H_{\rm L}$ and the orthogonal linear
combination which is heavy by $H_{\rm H}$.

\begin{eqnarray}
H_{\rm L} &=& {\rm sin} \; \beta \; H_u \; + \; {\rm cos} \; \beta \; H_d^* \\
H_{\rm H} &=& - \; {\rm cos} \; \beta \; H_u \; + \; {\rm sin} \; \beta \; H_d^*
\end{eqnarray}      
The potential for $H_{\rm L}$ in the effective theory below the scale $\hat{m}$ has 
the simple form:
\begin{equation}
V_{\rm L} = m_{\rm L}^2 |H_{\rm L}|^2 + 
\frac{{g_2}^2 + {g_1}^2}{8} \; {\rm cos}^2 \; 2\beta \; |H_{\rm L}|^4 
\end{equation}
Here the parameter $m_{\rm L}^2$ is given by:
\begin{equation}
m_{\rm L}^2 = \left( 2 \delta{\mu} \; \hat{\mu} + \delta{\mu}^2   
+ {{\delta{m}}_u}^2 \; {\rm sin}^2 \; \beta + 
{{\delta{m}}_d}^2 \;
{\rm cos}^2 \; \beta \right) 
+ \; \delta{B} \; {\rm sin} \; 2 \beta   
\end{equation}
The condition for electroweak symmetry breaking is that $m_{\rm L}^2 < 0$.
Now if $\hat{\mu}$ is not very large and if tan$^2 \; \beta \gg 1$ this is 
equivalent to requiring $\delta m_u^2 < 0$. Therefore the criterion for
getting the right pattern of symmetry breaking is that ${\hat{m}^2}_u < 0$,
$\delta m_u^2 < 0$ with $|{\hat{m}^2}_u | \gg | \delta m_u^2|$. 

Since the quartic terms in the potential for $H_{\rm L}$ depend on the gauge
coupling strengths of SU(2$)_{\rm L}$ and U(1$)_{\rm Y}$ we can determine the mass
of the lightest neutral Higgs at tree level to be $m_{\rm Z}$ cos 2$\beta$. However
just as in the MSSM the quartic term in the Higgs potential receives significant
corrections from loops involving the top Yukawa coupling. The relevant part of the
superpotential is the SU(3) invariant top Yukawa coupling, which we denote by
$\lambda$ and the mass term that decouples the extra colored states.
\begin{equation}
\int \; d^2 \theta  \; \lambda \left( \; Q_3 H_u t^c + \overline{T^c} S_u t^c
\; \right) + \; M \; T^c \; \overline{T^c} 
\end{equation}
We are particularly interested in the limit $M > \lambda \vev{S_u}$ because the 
physical top Yukawa is given by
\begin{equation}
\lambda_{t} = \lambda \; \frac{M}{\sqrt{M^2 \; + \; \lambda^2  \vev{S_u}^2}}
\end{equation}
This implies that in order to get a physical top Yukawa of order one we 
typically require $M > \lambda \vev{S_u}$. The bound on the mass of the 
extra U(1) gauge boson typically requires that $\vev{S_u} \geq 2.5$ TeV
{\cite{Bogdan}}, which then implies
that in the region of interest $\vev{S_u} > \hat{m}$ as well. Therefore we have 
the hierarchy $M > \lambda \vev{S_u} > \hat{m}$.  For the effective theory 
below the scale $M$, the relevant part of the superpotential is then exactly
as in the MSSM.
\begin{equation}
\int \; d^2 \theta \; \lambda_{t} \; Q_3 H_u t^c   
\end{equation}
This gives the familiar logarithmically enhanced
one-loop contribution to the Higgs quartic coupling
\begin{equation}
\Delta \; V_{\rm B} = \frac { 3 \lambda_t ^4}{ 16 \pi^2} |H_u|^4 {\rm ln}
\left(\frac{m_{\tilde{t}}^2}{m_{t}^2}\right)
\end{equation} 
The effect of this, as in the MSSM, is to raise the mass of the lightest neutral
Higgs
above M$_{\rm Z}$.

\subsection{Radiative Stability of the Global Symmetry}

We now investigate the logarithmically enhanced radiative corrections to the Higgs
potential with a view to understanding whether they give rise to the pattern of
symmetry breaking discussed in the previous section, and whether the global
symmetry is stable under quantum corrections. For simplicity we ignore all
superpotential couplings in the analysis apart from the top Yukawa and the $\mu$
terms. We also ignore other small effects.

We parametrize the superpotential as
\begin{equation}
\int d^2 \theta \; \left( \lambda_u Q_3 H_u t^c + \lambda_s \overline{T^c} S_u t^c
+ \hat{\mu} \; S_u S_d \; + {\mu} H_u H_d \right)
\end{equation}
At some ultraviolet scale $\Lambda$ the SU(3) global symmetry is exact, which
implies $\lambda_u = \lambda_s$ and $\hat{\mu} = \mu$. However radiative corrections
from scales below $\Lambda$ will in general alter these relations. The 
renormalization
group equations for the Yukawa couplings are
\begin{eqnarray}
\frac{ d \; \lambda_u}{ d \; t} &=& \frac{1}{16 \pi^2} \lambda_u 
\left(6 {\lambda_u}^2 + {\lambda_s}^2 - \frac{16}{3} {g_3}^2 - 3 {g_2}^2 - 
\frac{13}{9} {g_1}^2 \right)  \\
\frac{ d \; \lambda_s}{ d \; t} &=& \frac{1}{16 \pi^2} \lambda_s
\left(2 {\lambda_u}^2 + 5 {\lambda_s}^2 - \frac{16}{3} {g_3}^2 
- \frac{16}{9} {g_1}^2 \right) .
\end{eqnarray}
We see that these couplings receive corrections from SU(2$)_{\rm L} \times$ 
U(1$)_{\rm Y}$ 
gauge
loops that violate the SU(3) global symmetry, so that at low scales we do not expect
them to have the same value. Close to the weak scale we expect that each coupling will
be driven to a fixed point value just as in the MSSM; however the two fixed point
values will now no longer be the same. The difference between the fixed point values 
is given by
\begin{equation}
\label{eq:difference}
\lambda_u^2 - \lambda_s^2 = \frac{1}{4} \left(3 g_2^2 - \frac{1}{3} g_1^2 \right) .
\end{equation}
Since this difference is small compared to the individual couplings themselves 
global SU(3) is still a good approximate symmetry of the Yukawa couplings
at low energies.

What about the renormalization group equations for $\hat{\mu}$ and $\mu$? These have
the form:
\begin{eqnarray}
\frac{ d \; \hat{\mu}}{ d \; t} &=& \frac{1}{16 \pi^2} \hat{\mu}
\left(3 {\lambda_s}^2 \right)  \\
\frac{ d \; \mu}{ d \; t} &=& \frac{1}{16 \pi^2} \mu
\left(3 {\lambda_u}^2 - 3 {g_2}^2 - {g_1}^2 \right)
\end{eqnarray}
In contrast to the Yukawa couplings we see that the difference $\delta{\mu}$ between
$\hat{\mu}$ and $\mu$ at low energies, although loop suppressed is logarithmically
enhanced, and can be comparable to the absolute values of $\hat{\mu}$ and $\mu$
themselves. We conclude that global SU(3) is typically not a good approximate
symmetry of the dimensionful mass terms for the Higgs fields in the superpotential.
We therefore require that these terms be small compared to $\hat{m}$ in order that
the approximate global symmetry of the Higgs potential $V_{\rm tot}$ is preserved.

We now consider the dominant terms in the 
renormalization group equations for the soft masses of $H_u$ and
$S_u$:
\begin{eqnarray}
\frac{ d \; m^2_{s}}{ d \; t} &=& \frac{1}{16 \pi^2} \left[
3 \lambda_{s}^2 \left( 2 m^2_{\tilde{t}} + 2 m^2_{s} + 2 m^2_{\overline{T}} \right)
\right] \\
\frac{ d \; m^2_{u}}{ d \; t} &=& \frac{1}{16 \pi^2} \left[
3 \lambda_{u}^2 \left( 2 m^2_{\tilde{t}} + 2 m^2_{u} + 2 m^2_{\tilde{Q}} \right)
- 6 g_2^2 \; M_2^2 \; - 2 g_1^2 \; M_1^2 \right]
\end{eqnarray}
In these equations $M_2$ and $M_1$ are the masses of the SU(2) and U(1) gauginos. We
see from this that $m^2_{s}$ and $m^2_{u}$ are driven negative so that radiative
breaking of the SU(3) global symmetry occurs exactly as in the MSSM through loops
involving the top Yukawa coupling. As shown in the previous subsection the VEV $f$
naturally lies along $S_u$ and $S_d$, the electroweak singlet directions, because of
the D terms of the SU(2$)_{\rm L} \times$ U(1$)_{\rm Y}$ gauge group.

What about electroweak symmetry breaking? Does it also naturally occur? As
explained in the previous subsection the criterion for this is that $ \delta m_u^2
= \; m^2_{u} - m^2_{s} < 0$. The one loop renormalization group equations for the
soft masses above would seem to indicate that $\delta{m_u}^2$ is in fact greater
than zero, due to the net positive contribution from gauge and gaugino loops to
the mass of $H_u$. However, a closer inspection reveals that this is not the case.
Since at low scales $\lambda_u$ and $\lambda_s$ differ by the amount shown in
({\ref{eq:difference}}), and in particular $\lambda_u > \lambda_s$ we see that
there is a net two loop contribution to the mass splitting that drives
$\delta{m_u}^2 < 0$. If the stop masses are larger than the SU(2$)_{\rm L}$ and
U(1$)_{\rm Y}$ gaugino masses the two loop contribution is dominant so that
electroweak symmetry breaking can take place. This is indeed the case for most
realistic supersymmetric spectra.

In these models there are restrictions from naturalness on the scale at which the
soft superpartner masses are generated. We denote this scale by $\Lambda_{\rm 
soft}$. Then the renormalization group equations imply that at low scales 
\begin{equation}
|\delta m_u^2| \approx \frac{3} {4 \pi^2} \left( \lambda_u^2 - \; \lambda_s^2 
\right) \; \hat{m}^2 \; {\rm log}\left(\frac{\Lambda_{\rm soft}}{\hat{m}} \right)
\approx 
\frac{\hat{m}^2}{50} \; {\rm log}\left(\frac{\Lambda_{\rm
soft}}{\hat{m}} \right) .
\end{equation}
If the logarithm above becomes of order 10 the hierarchy between $\delta{m}$ and
$\hat{m}$ is affected. This shows that in this scenario low values of the scale
$\Lambda_{\rm soft}$ are preferred by naturalness. This 
class of theories includes models of low 
energy
supersymmetry breaking such as gauge mediation {\cite{GMSB}}, 
where $\Lambda_{\rm soft}$ is
the scale of the messenger masses. 

The fact that the Standard Model matter fields are charged under U(1)$_{\rm E}$
also puts important restrictions on the soft scalar masses. The reason is that the
non-zero D-term of U(1)$_{\rm E}$ will in general give a correction of order
$\hat{m}^2$ to the soft mass squared of each scalar superpartner, the sign of 
which
may be positive or negative, depending on the charge of the field under U(1$)_{\rm
E}$. We therefore require that those fields which get a negative correction from
the D-term of U(1)$_{\rm E}$ receive a larger positive contribution to their soft
masses from supersymmetry breaking. This places restrictions on the manner in
which supersymmetry is mediated to the visible sector fields. In the absence of
fine-tuning this generally implies that all the soft scalar masses are at least of
order $\hat{m}$, except the physical Higgs.

The constraint that the $\mu$ terms be small, in conjunction with the constraints
({\ref{stability}}) on the parameters of the Higgs potential also places
restrictions on the manner in which supersymmetry breaking is communicated to the
visible sector. In particular there should not be too much of a hierarchy between
the stop masses and the up and down Higgs masses at $\Lambda_{\rm soft}$, so that
even when electroweak symmetry is broken the sum of the up-type and down-type soft
Higgs mass squared remains positive.

\section{Grand Unification}

\subsection{Embedding of the Gauge and Global Symmetries} 

We now explain how the framework we have outlined in the previous section can be
successfully incorporated into a supersymmetric grand unified theory, thereby
maintaining the prediction of coupling constant unification. The approach we will
take is to identify the global symmetry with a subgroup of a larger grand unifying
group, which we take to be SU(6). The reason for this choice is that since SU(6)
contains SU(3$) \times$ SU(3$) \times$ U(1) as maximal subgroups it is natural to
identify one of these two SU(3) groups with color and the other with the global
symmetry we require the theory to possess. But why then is one of these two SU(3)
groups completely gauged while only an SU(2$)_{\rm L} \times$ U(1$)_{\rm Y}$
subgroup of the remaining SU(3$) \times$ U(1) is gauged? This can be understood as a
consequence of the pattern of symmetry breaking. We consider a scenario where there
are two separate effects which each break SU(6), but in two completely 
different breaking
patterns: \\
\\
The first breaks $\; \; \; \; \; \; \; \; \; $ 
SU(6$) \rightarrow$ SU(5) \\
The other breaks  $\; \; \; \; \; \; \; $
SU(6$) \rightarrow$ SU(3$) \times$ SU(3$) \times$ U(1). \\
\\
Their net result is $\; \; \; \; \; \; $
SU(6) $\rightarrow$ SU(3$) \times$ SU(2$)_{\rm L} \times$ U(1$)_{\rm Y}$. \\ 
\\
For example the first pattern can be obtained 
if the SU(5) singlet components of a 6, $\bar{6}$ pair acquire VEVs
, while the second can
be obtained if an adjoint of SU(6), a $35$, acquires a VEV along 
Diag$\left(1,1,1,-1,-1,-1 \right)$. 
We denote these VEVs by
$\vev{6}, \vev{\bar{6}}$ and $\vev{35}$ respectively. For now we will
assume that this pattern of VEVs is responsible for the breaking of SU(6), 
although our results are in fact more general.  
If the Yukawa coupling of the top quark is screened from the first 
of the two effects above that
break SU(6) it will be invariant under the SU(3) global 
symmetry up to loop corrections. This is what we require for our scenario to
be viable.  

How can the Standard Model quarks, leptons and Higgs fields be embedded in the
SU(6) group? The SU(6) representations we require are the $6$ and $\bar{6}$, the
15, which is the two index antisymmetric tensor, and the 20, which is the three
index antisymmetric tensor. Under the SU(5) subgroup of SU(6) these representations
decompose in the following way:
\begin{eqnarray}
6 &\rightarrow& 5 + 1 \\
\bar{6} &\rightarrow& \bar{5} + 1 \\
15 &\rightarrow& 5 + 10 \\
20 &\rightarrow& 10 + \overline{10} \\
35 &\rightarrow& 24 + 5 + \overline{5} + 1 
\end{eqnarray}
In an SU(5) grand unified theory one Standard Model family of quarks and leptons
emerges from the $\bar{5}$ and 10 representations. The simplest generalization of
this to SU(6) involves two $\bar{6}$ representations and a 15; this particular
combination is anomaly free. Once SU(6) is broken down to SU(5) the 5 from the 15 
and one of two $\bar{5}$'s from the 
$\bar{6}$'s are 
now vector-like under the
unbroken group and can be given a mass and decoupled, along with the singlets,
leaving the familiar chiral $\bar{5}$ and 10 at low energies. The interactions 
that do this are of the form 
\begin{equation}
\label{trouble}
\vev{\bar{6}} 15 \; \bar{6} \; \; , \; \frac{\vev{6}\vev{6}}{\Lambda} \; \bar{6} 
\; \bar{6}
\end{equation}
Here $\Lambda$ is an ultraviolet scale, which may either be the Planck scale or
a scale associated with the mass of a singlet that is integrated out to obtain this
operator.   

What about the Higgs fields?
In an SU(5) grand unified
theory the up and down Higgs fields emerge from the doublets in the 5 and $\bar{5}$
representations respectively. In the most straightforward generalization of this to
SU(6) the Higgs fields emerge from the SU(2) doublets in a $6$, $\bar{6}$ pair. 
This
however leads to a difficulty. While the down-type Yukawa couplings can be easily
generalized from SU(5) to SU(6) 
\begin{equation}
\label{downYukawa}
\bar{5}_{\bar{{\rm H}}} \; 10 \; \bar{5} \rightarrow \bar{6}_{\bar{{\rm H}}} 
\; 15 \; \bar{6} 
\end{equation}
there is no way to write an SU(6) invariant up-type Yukawa coupling with this 
minimal choice of matter and Higgs representations. However once SU(6) is broken 
to SU(5) it is possible to write such couplings.
For example, 
the familiar up-type Yukawa 
coupling of SU(5) can be generalized to SU(6)
\begin{equation}
\label{up}
{5}_{{\rm H}} \; 10 \; 10 \rightarrow \frac{\vev{6}}{\Lambda} \; {6}_{{\rm H}} \; 15 
\; 15 .
\end{equation}
Here $\Lambda$ is either the Planck scale or a scale associated with the
mass of a vector-like $15, \overline{15}$ pair that is integrated out to obtain
this operator {\cite{CLP}}. While this is satisfactory for the first two
generations it does not serve our purposes as regards the third generation.  
In particular, since this operator breaks SU(6), an up-type mass term of 
this 
form does not preserve the SU(3) global 
symmetry that we wish the top Yukawa to respect.  
While there may be more than one possible solution to this problem 
the approach we will follow here is to introduce into the theory a 20 
dimensional representation of SU(6), which is a three-index antisymmetric tensor. 
Then the up-type Yukawa 
coupling of 
SU(5) can be generalized to SU(6) as {\cite{Berezhiani}}
\begin{equation}
\label{top61520}
{5}_{{\rm H}} \; 10 \; 10 \rightarrow {6}_{{\rm H}} \; 15 \; 20 .
\end{equation}
Since this interaction is SU(6) invariant it naturally respects the SU(3) 
global symmetry. However, the model now contains an extra $10$ and $\overline{10}$ 
of SU(5) beyond the fields in the MSSM.

In conventional grand unified models based on the SU(6) group once SU(6) is broken
to SU(5) or its subgroups the unwanted $\overline{10}$ (of SU(5)) coming from the 20
pairs up with one linear combination of the 10's from the 15 and from the 20,
acquires a mass and decouples. Therefore the physical third generation fields
emerge from the orthogonal linear combination. The interactions that lead to this
have the form
\begin{equation}
\label{boot}
\int d^2 \theta \; \vev{6} 15 \; 20 + 20 \vev{35} 20 .
\end{equation}
The net result is that at low energies 
we are left with precisely the familiar 3 
generations which have the quantum numbers of $\bar{5}$ and 10 under SU(5).
This however does not serve our purpose, since decoupling the extra 10 
and $\overline{10}$ destroys the global SU(3) symmetry of the top Yukawa coupling 
that we are trying to preserve. We therefore assume that such terms are forbidden 
until supersymmetry is broken, in analogy with the $\mu$ term of the MSSM, and 
that instead we have
\begin{equation}
\label{weakmass}
\int d^2 \theta \; \tilde{\mu} \; \frac{\vev{35}}{\Lambda} \; 20 \;20 \; ,
\; \; \; \;
\int d^2 \theta \; \tilde{\mu}  \; \frac{\vev{6}}{\Lambda} \; 15 \; 20 
\end{equation}
where $\tilde{\mu}$ is of order $\hat{m}$, and $\Lambda$
is an ultraviolet scale which we assume is close to (but above) the unification
scale. Then the additional $10$ and $\overline{10}$ are present in the theory until
the weak scale, and therefore the global SU(3) symmetry of the top Yukawa is
maintained until then.

Therefore to summarize, the matter content of this class of unified models is given
by 3 generations of $\left[\; \bar{6}, \bar{6}, 15 \; \right]$ and a single 20,
while the light Higgs fields emerge from a 6 and a $\bar{6}$.  A subset of this
matter content, constituting three generations of fields which transform as 5,
$\bar{5}$ under SU(5), along with six SU(5) singlets (two per generation), is
vector-like under the Standard Model gauge group and decouples near the
unification scale. However, another subset of this matter content, which has the
quantum numbers of a 10 and $\overline{10}$ under SU(5), is also vector-like under
the Standard Model gauge group but remains light until the weak scale thereby
preserving the global symmetry of the top Yukawa coupling.
 
In this scenario the pseudo-Goldstone nature of the physical Higgs can be
understood as a consequence of the fact that the VEVs of 6$_{\rm H}$ and
$\bar{6}_{\bar{\rm H}}$ are aligned along the same direction as $\vev{6}$ and
$\vev{\bar{6}}$, while at the same time the potentials of $6_{\rm H}$ and
$6_{\bar{\rm H}}$ are screened from $\vev{6}$ and $\vev{\bar{6}}$. Therefore an
SU(3) subgroup of the SU(6) gauge symmetry survives as an approximate global
symmetry of the sector of the theory containing $\hat{H}_u$ and $\hat{H}_d$. This
idea is similar in spirit to that of the little Higgs model of Kaplan and Schmaltz
{\cite{KS}}. A major difference is the large hierarchy of scales between the two
pairs of VEVs which makes logarithmically enhanced radiative corrections to the
potential for the light Higgs fields important.

\subsection{Realization of the Global Symmetry}

We now study in more detail how the global symmetry is realized in this class of
models, and in particular how the fermion masses emerge.  For this purpose it is
useful to understand how the various SU(6) representations decompose under the
SU(3) $\times$ SU(3) $\times$ U(1) subgroup of SU(6). For simplicity, in what 
follows we
suppress the U(1) quantum numbers, except to distinguish between the two singlets
of SU(3)$\times$SU(3) in the 20. While one of these has the U(1) quantum numbers
of $\tau^c$ the other has the quantum numbers of the complex conjugate of 
$\tau^c$.
\begin{eqnarray}
6 &\rightarrow& (3,1) + (1,3) \\
\bar{6} &\rightarrow& (\bar{3},1) + (1,\bar{3})\\
15 &\rightarrow& (3,3) + (\bar{3},1) + (1, \bar{3}) \\
20 &\rightarrow& (3, \bar{3}) + (\bar{3},3) + (1,1)_{\tau^c} + 
(1,1)_{\overline{\tau}} \\
35 &\rightarrow& (8,1) + (3, \bar{3}) + (\bar{3},3) + (1,8) + (1,1)
\end{eqnarray}

Consider first the top Yukawa coupling ({\ref{top61520}). Under SU(3)
$\times$ SU(3) this decomposes as shown below
\begin{eqnarray}
\label{top3}
6_{\rm H} \; 15 \; 20 \rightarrow \left\{ \begin{array}{l} (1,3)_{\rm H} (\bar{3},1) (3, \bar{3}) \\
(1,3)_{\rm H} (3,3) (\bar{3},3) \\
(1,3)_{\rm H} (1,\bar{3}) (1,1)_{\bar{\tau}} \end{array} \right.
\end{eqnarray}
where we have ignored the interactions of the color triplet Higgs, which are 
irrelevant to our discussion.
The first of the three terms on the right hand side above is precisely the
SU(3) invariant interaction ({\ref{top}}) that we require. We must ensure that the 
form of this term is not disturbed by any SU(3) violating interaction except at 
loop level.

Consider now the interaction ({\ref{trouble}}) which decouples the vector-like
states in the $\bar{6}$'s and the 15 at the unification scale. Under SU(3)
$\times$ SU(3) it decomposes as shown below. 
\begin{eqnarray} 
\vev{\bar{6}} \; 15\; \bar{6} \rightarrow 
\left\{\begin{array}{l}\vev{1,\bar{3}} (3,3) (\bar{3},1),\\
\vev{ 1,\bar{3}} (1,\bar{3}) (1,\bar{3}) \end{array} 
\right. 
\end{eqnarray}
Although this interaction removes states which are not in complete multiplets of
the global SU(3), notice that it does not directly affect any of the states in the
SU(3) invariant top Yukawa interaction, the first line on the right hand side of
({\ref{top3}}). However it does affect the interactions on the second and third
lines on the right hand side of ({\ref{top3}}), which are no longer SU(3$) \times$
SU(3) invariant in the low energy effective theory, but instead transform as shown
below under SU(3$) \times$ SU(2$)_{\rm L}$ of the Standard Model.
\begin{eqnarray}
\label{horrorone}
(1,3)_{\rm H} (3,3) (\bar{3},3) &\rightarrow& 
[1,{2}]_{\rm H}[3,{2}][\bar{3},{1}]  
+ [1,{1}]_{\rm H}[3,{2}][\bar{3},{2}] \\
\label{horrortwo}
(1,3)_{\rm H} (1,\bar{3}) [1,1]_{\bar{\tau}}
&\rightarrow& [1,{1}]_{\rm H}[1,{1}][1,{1}]_{\bar{\tau}}
\end{eqnarray}
Above and in what follows we use ordinary brackets to denote quantum numbers under
SU(3)$\times$ SU(3) and square brackets to denote quantum numbers under SU(3$)
\times$ SU(2$)_{\rm L}$. Since the interactions ({\ref{horrorone}})
and ({\ref{horrortwo}}) violate the 
global SU(3) we
require that they be small so that they do not significantly feed into the Higgs
soft mass at loop level.  However the SU(6) symmetry of the operator
({\ref{top61520}) that generates all these couplings would imply that 
all of the couplings in ({\ref{top3}}) have the same strength. We therefore
require that there be significant 
contributions that violate SU(6) but respect SU(3$) \times $ SU(3)
in the sector that generates the top Yukawa. These could arise from operators
such as the one below.
\begin{eqnarray}
\frac{\vev{35}}{\Lambda} 6 \; 15 \; 20 &\rightarrow& \left\{\begin{array}{l}(1,3) (\bar{3},1) (3, \bar{3}), \\
 (1,3) (3,3) (\bar{3},3), \\
 (1,3) (1,\bar{3}) (1,1)_{\bar{\tau}} \end{array}\right.
\end{eqnarray}
We see that this decomposes under SU(3$) \times$ SU(3) into the very same operators as in
({\ref{top3}), but these now have different coefficients.

We now move on to consider the operators ({\ref{weakmass}}), which give masses
of order $\hat{m}$ for the remaining light states which are vector-like under SU(5)
\begin{eqnarray}
\label{mass10}
\int d^2 \theta \; \tilde{\mu} \; \frac{\vev{35}}{\Lambda} \; 20 \;20
&\rightarrow&  \left\{\begin{array}{l}\tilde{\mu} \; (3, \bar{3}) (\bar{3}, 3), \\
\tilde{\mu} \; (1,1)_{\bar{\tau}} (1,1)_{{\tau}^c} \end{array}\right.
\end{eqnarray}

\begin{eqnarray}
\label{mass11}
\int d^2 \theta \; \tilde{\mu} \; \frac{\vev{6}}{\Lambda} \; 15 \; 20 
&\rightarrow & \left\{\begin{array}{l}\tilde{\mu} \; [\bar{3},1] [3,1], \\
\tilde{\mu} \; [3,2] [\bar{3},2], \\
\tilde{\mu} \; [1,1] [1,1]_{\bar{\tau}} \end{array}\right.
\end{eqnarray}
These masses are not invariant under the global SU(3). When $S_u$ (which 
in our notation is
$[1,1]_{\rm H}$) acquires a VEV we see from the form of the interactions 
(\ref{top3}), 
(\ref{mass10}) and (\ref{mass11})
that one linear combination of the $[\bar{3}, {1}]$
from the 15 and the $[\bar{3},{1}]$ from the 20 gets a mass with the $[{3}, {1}]$
from the 20 leaving the orthogonal linear combination light. This light state is the
physical $t^c$. It is straightforward to establish that from
(\ref{horrorone}), (\ref{horrortwo}), (\ref{mass10}) and (\ref{mass11}) that the
physical $Q_3$ and $\tau^c$ are also linear combinations of states from the 15 and
the 20.

We now come to the third generation Yukawa couplings in the down sector. These 
emerge from the
operator ({\ref{downYukawa}}), which decomposes under SU(3$) \times$ SU(3)
as shown below 
\begin{eqnarray}
\bar{6}_{\bar{\rm H}} \; 15 \; \bar{6} &\rightarrow& \left\{\begin{array}{l}(1, \bar{3})_{\bar{\rm H}} (3,3) 
(\bar{3},1), \\
(1, \bar{3})_{\bar{\rm H}} (1,\bar{3}) (1, \bar{3}) \end{array}\right.
\end{eqnarray}
As a consequence of the operator (\ref{trouble}) which decouples states that are 
not in 
complete 
representations of the global SU(3), these interactions are not invariant under
the global symmetry below the unification scale. At low energies they have their
familiar Standard Model forms
under SU(3)$\times$SU(2$)_{\rm L}$ as shown below.
\begin{eqnarray}
\label{bottom}
(1, \bar{3})_{\bar{\rm H}} (3,3) (\bar{3},1)  &\rightarrow& 
[1,{2}]_{\bar{\rm H}}
[3,{2}] [\bar{3},{1}] \\
\label{tau} 
(1, \bar{3})_{\bar{\rm H}} (1,\bar{3}) (1, \bar{3}) &\rightarrow& 
[1,{2}]_{\bar{\rm H}} [1,{1}]_{\tau^c}
[{1},{2}]
\end{eqnarray}
Although these couplings are not invariant under the global symmetry they are not
large enough to significantly affect the SU(3) symmetric form of the top Yukawa.

What about the first two generations? How do their Yukawa couplings arise? The
down-type masses arise from operators of the very same form ({\ref{downYukawa}}) as
for the third generation. At the unification scale all the states in these two
generations which are vector-like or singlets under the Standard Model gauge group
get a mass from the operator (\ref{trouble}), so that at low energies only the
Standard Model fields survive. The down-type interactions then take their familiar
Standard Model form, exactly as in (\ref{bottom})  and (\ref{tau}) above. The up-type
Yukawa couplings however, in contrast to the top Yukawa, arise from the operator
(\ref{up}). This operator manifestly breaks the global SU(3) symmetry at the
unification scale. However it is too small to significantly affect the SU(3)
symmetric form of the top Yukawa coupling. It decomposes under SU(3) $\times$
SU(2$)_{\rm L}$ precisely into the up-type Yukawa couplings of the Standard Model.
\begin{equation}
\frac{\langle 6\rangle}{\Lambda} \; 6_{\rm H} \; 15 \; 15 \rightarrow 
[1,{2}]_{{\rm H}} 
[3,{2}] [\bar{3},{1}]
\end{equation}

\subsection{A Realistic Unified Model}

We now construct a realistic unified model based on the ideas we have outlined. In
order to solve the doublet-triplet splitting problem as in {\cite{Kawamura}} we
consider models with a compact fifth dimension where the breaking of SU(6)
$\rightarrow$ SU(3) $\times$ SU(3) $\times$ U(1) is realized by imposing
appropriate boundary conditions on fields. The extra dimension is assumed to be
extremely small, with size of order the unification scale, which is about
$10^{15} - 10^{16}$ GeV. In contrast however, the breaking of SU(6) to SU(5) is 
realized by
the expectation values $\vev{6}$ and $\vev{\overline{6}}$ of fields transforming
in the fundamental and anti-fundamental representations of SU(6). Provided that
this theory is strongly coupled at the cutoff of the higher dimensional theory, we
expect the coupling constants to unify at about the same level as in the MSSM
{\cite{HN}}.

We now consider the breaking of SU(6) $\rightarrow$ SU(3) $\times$ SU(3) $\times$
U(1) in more detail. Explicitly, we start with a five dimensional SU(6) gauge
theory compactified on an $S^1/Z_2$ orbifold. Compactification on $S^1/Z_2$ is
obtained by identifying the fifth coordinate $x_5$ under the two operations $x_5
\rightarrow -x_5$ (reflection) and $x_5 \rightarrow x_5 + 2 \pi R$ (translation).  
Then the physical space is the line interval between $x_5 = 0$ and $x_5 = \pi R$.
We assume that
there are branes at the orbifold fixed points $x_5 = 0$ and $x_5 = \pi R$ where 
fields can be localized. The
boundary conditions are chosen so that the reflection symmetry breaks 5D N = 1
supersymmetry to 4D N = 1 supersymmetry while the translation symmetry breaks
the gauge symmetry. Specifically we start with a five dimensional SU(6) gauge
field 
$A_M \equiv A_M^a
T^a, M=1 \ldots 5, a=1\ldots 35$, and impose the following boundary
conditions:
\begin{eqnarray}
\label{bcs}
A_\mu (x^\mu, x_5) &=& +A_\mu (x^\mu, -x_5) = Z\,A_\mu (x^\mu, x_5+2\pi R)\,
Z^{-1}, \nonumber  \\
A_5 (x^\mu, x_5) &=& -A_5 (x^\mu, -x_5) = Z\,A_5 (x^\mu, x_5+2\pi R)\,
Z^{-1},
\end{eqnarray}
where $\mu=1\ldots 4$ and $Z=$diag$(+,+,+,-,-,-)$. The low-energy effective field
theory in four dimensions contains the seventeen massless gauge bosons $A_\mu$ of
the unbroken SU(3) $\times$ SU(3) $\times$ U(1) group, while the remaining eighteen
fields $X_\mu$ which also transform as vector fields in four dimensions do not have
massless modes. Similarly the thirty five fields $A_5^a$ which transform as four
dimensional scalars do not possess zero modes. Since we are working in the context
of a supersymmetric model appropriate boundary conditions must also be imposed on
the other fields in the higher dimensional gauge supermultiplet. What are these? In
addition to the five dimensional vector field $A_M$, a five dimensional gauge
supermultiplet consists of a symplectic Majorana spinor $\lambda_i$ and a real
scalar $\sigma$. The boundary conditions for these fields are given by the same
equation ({\ref{bcs}}), with $\lambda_{1+} = \frac{1}{2} \left( 1 + \gamma_5
\right) \lambda_1$ transforming like $A_{\mu}$ while $\sigma$ and $\lambda_{2+} =
\frac{1}{2} \left( 1 + \gamma_5 \right) \lambda_2$ transform like $A_5$. The zero
modes of $A_{\mu}$ and $\lambda_{1+}$ form a 4D $N = 1$ gauge multiplet, while the
other fields have no massless modes.

In such a framework supersymmetric chiral multiplets in the four dimensional
effective theory can emerge in two ways - from five dimensional hypermultiplets
propagating in the bulk of the space or from four dimensional chiral multiplets
constrained to the boundaries of the five dimensional space. Consider first a five
dimensional hypermultiplet $\Psi$ transforming in the fundamental of SU(6). $\Psi$
consists of a Dirac fermion $\psi$ and two complex scalars $\phi$ and $\phi^c$. The
boundary conditions on the scalars are,
\begin{eqnarray}
\phi (x^\mu, x_5) &=& +\phi (x^\mu, -x_5) = {\cal C}\,Z\,\phi (x^\mu, x_5+
2\pi R),\\
\phi^c (x^\mu, x_5) &=& -\phi^c (x^\mu, -x_5) = {\cal C}\,Z\,\phi^c (x^\mu,
x_5+2\pi R),
\end{eqnarray}
where as before $Z=~$diag$(1,1,1,-1,-1,-1)$, and ${\cal C}=\pm 1$ is the parity of
the field $\Psi$. Now under SU(3) $\times$ SU(3) the components of $\phi$ and
$\phi^c$ transform as $\left[(3,1),(1,3)\right]$. As a consequence of these
boundary conditions $\phi^c$ does not have any massless mode at all, while between
the (3,1) component and the (1,3) component of $\phi$ only one has a massless mode
while the other does not. The fields $\psi_{+} = \frac{1}{2}(1 + \gamma_5)\psi$ and
$\psi_{-} = \frac{1}{2} (1 - \gamma_5)\psi$ obey the same boundary conditions as
$\phi$ and $\phi^c$, respectively. Therefore $\psi_{-}$, like $\phi^c$ has no 
massless modes. The fields $\phi$ and $\psi_{+}$ each have a 
massless mode which
together form a four dimensional $N=1$ chiral multiplet that transforms either as
(3,1) {\bf or} as (1,3) under the unbroken SU(3) $\times$ SU(3) gauge group.
What about bulk hypermultiplets transforming under other representations of SU(6)?
These also lead in general to massless chiral multiplets which do not transform as
complete representations of SU(6) but only as complete representations of the
unbroken subgroup.

We now consider the case where chiral multiplets emerge from fields localized to
the boundaries of the space at $y = 0$ or $y = \pi R$. The point $y = \pi R$ is
called the `3-3-1' point since the wave functions of all the $X_{\mu}$ gauge bosons
vanish there. Fields localized to this point need not necessarily be in complete
representations of SU(6) but need only be in complete representations of the
unbroken group SU(3) $\times$ SU(3) $\times$ U(1). Similarly interactions localized
to this point need only be invariant under the unbroken group. However fields
localized to the point $y = 0$ must be in complete representations of SU(6), and
interactions at this point must also be invariant under the full SU(6) gauge 
symmetry.

We are now in a position to explain the matter content of the model. It consists of
three generations of fields transforming under SU(6) as
\begin{equation}
\begin{array}{cccccc}
\bar{6}_{-1} & \bar{6}_{-3} & 15_1 & 1_5 & 1_3 & 1_1 
\end{array}
\end{equation}
where the subscripts give the charges of the various fields under the additional
U(1$)_{\rm E}$. In addition to this matter content there are two fields 
each
transforming under SU(6) as the 20 dimensional representation. They have charges +1
and -1 under U(1$)_{\rm E}$. It is straightforward to verify that matter content
with this set of charge assignments is free of any anomalies. 
  
We now turn to the Higgs content. It consists of two fields in the fundamental and
anti-fundamental of SU(6) which break SU(6) to SU(5), and which are uncharged under
U(1$)_{\rm E}$. The VEVs of these fields are denoted by $\vev{6}_0$ and
$\vev{\bar{6}}_0$. The Higgs fields responsible for the breaking of the global
SU(3) symmetry and electroweak symmetry also emerge from the fundamental and 
antifundamental representations of SU(6), and we denote these fields by $6_{\rm H}$ 
and $\bar{6}_{\bar{\rm H}}$. Their charges under U(1$)_{\rm E}$ are -2 and +2 
respectively. It is clear from this that the Higgs content is also anomaly free.

We now consider the locations of the various matter fields in the higher
dimensional space. Both fields transforming as the 20 of SU(6) are 
hypermultiplets propagating in the
bulk of the space. The boundary conditions project out some of the states 
in these representations so that
the massless fields in the low energy effective theory transform under 
SU(3) $\times$ SU(3) as shown below.
\begin{equation}
\begin{array}{cccc}
20_{+1} &\rightarrow& (1,1)_{+1, \bar{\tau}} & (3,\bar{3})_{+1} \\
20_{-1} &\rightarrow& (1,1)_{-1, {\tau}^c} & (\bar{3},3)_{-1}
\end{array}
\end{equation}
The three generations of matter fields are all localized on the brane at $y = 0$.
This gives us an understanding of their Standard Model quantum numbers and 
the quantization of electric charge. 

What about the Higgs fields? Where are they located in the extra dimension? We
assume that the fields that break SU(6) to SU(5) propagate in the bulk of the
higher dimensional space while the fields that break electroweak symmetry are
localized to the brane at $y = 0$. Doublet-triplet splitting is realized by having
a fundamental and anti-fundamental of SU(6) in the bulk that have the same
U(1)$_{\rm E}$ quantum numbers as the electroweak Higgs fields on the brane. The
boundary conditions on the bulk fields are such as to leave the color triplets
light. A mass term between these bulk fields and the brane fields on the boundary
at $y = 0$ will then have the effect of giving mass to all the color triplets
while leaving one set of fields with the quantum numbers of $\hat{H}_u$ and
$\hat{H}_d$ light. These light fields are linear combinations of the original bulk
and brane Higgs fields.

We now turn our attention to the interactions of this theory. We start with the
top quark Yukawa coupling. The charges of the various fields under U(1$)_{\rm E}$ 
imply that of the two 20 dimensional representations only one can contribute to
the top Yukawa interaction below, which is localized at $y = \pi R$.
\begin{eqnarray}
6_{-2} \; 15_{1} \; 20_{1} &\rightarrow& \left\{\begin{array}{l}(1,3) (\bar{3},1) (3, \bar{3}), \\
 (1,3) (1,\bar{3}) (1,1)_{\bar{\tau}}\end{array}\right.
\label{moretau}
\end{eqnarray}
As before interactions of the form 
\begin{eqnarray}
\vev{\bar{6}}_{0} \; 15_{1} \; \bar{6}_{-1} &\rightarrow& \left\{\begin{array}{l}\vev{ 1,\bar{3} } (3,3) (\bar{3},1), \\
\vev{ 1,\bar{3} } (1,\bar{3}) (1,\bar{3})\end{array}\right.
\end{eqnarray}
decouple states in the 15 and the $\bar{6}$ which are vector-like under both 
SU(5) and U(1$)_{\rm E}$. This is true of matter in all three generations.
In addition there are now the interactions
\begin{equation}
\vev{\bar{6}}_{0} \bar{6}_{-1} 1_1 \; \; \;
\vev{\bar{6}}_{0} \bar{6}_{-3} 1_3
\end{equation}
whose effect is to decouple the SU(5) singlets which are vector-like under
U(1$)_{\rm E}$. 

We now consider the terms which give masses of order $\hat{m}$ to the light
fields in the two 20 dimensional representations. These take the form
\begin{equation}
\label{moreweakmass}
\tilde{\mu} \; 20_{+1} \; 20_{-1} \; \; , \; \; 
\tilde{\mu} \; \frac{\vev{6}_{0}}{\Lambda} \; 15_{1} \; 20_{-1}
\end{equation}
where $\tilde{\mu}$ is of order $\hat{m}$. The net result of the interactions
(\ref{moretau}) and (\ref{moreweakmass})
is that the physical $t^c$, $Q_3$ and $\tau^c$ are linear combinations of states in 
the 15 and the two 20's. 

All the down-type Yukawa couplings as well as the up-type Yukawa interactions for
the first two generations emerge from couplings of exactly the form discussed in
the previous subsection.  Constraints on the model arising from SU(5) mass
relations can be avoided if the matter fields on the brane with quantum numbers
$\bar{6}_3$ mix with additional fields in the bulk with the same quantum numbers,
so that the light fields are linear combinations of bulk and brane fields.
Alternatively, the first two generations of matter can emerge from bulk fields
rather than brane localized fields without significantly affecting any of the
other physics we have discussed.

It remains to consider the neutrino masses. How do they arise in this model? Perhaps
the simplest possibility is that the neutrinos acquire Dirac masses with the $1_5$
Standard Model singlet state. The smallness of the neutrino masses can be 
understood if a term of this form is forbidden until supersymmetry is broken, in
analogy with the $\mu$ term {\cite{Neal}}. 
\begin{equation}
\tilde{\mu} \; \frac{6_2}{\Lambda} \; 1_5 \; \bar{6}_{-3}
\rightarrow \frac{\tilde \mu}{\Lambda} [1,2]_{\rm H} [1,1] [1,2] 
\end{equation}
We conclude from this discussion that it is indeed possible to construct realistic 
little 
supersymmetric models in which the coupling constants unify.

\section{Conclusions}

We have proposed `little supersymmetry' as a solution to the supersymmetric little
hierarchy problem. This involves extending the Yukawa coupling of the top quark to
make it invariant under a global symmetry, of which the SU(2$)_{\rm L}$ subgroup is
gauged. The physical Higgs field is the pseudo-Goldstone boson associated with the
breaking of this global symmetry, and it can therefore naturally be significantly
lighter than the other scalar superpartners. We have shown that is possible to embed
the gauge and global symmetries together into a single grand unifying group, so that
the prediction of gauge coupling constant unification can be maintained.

These models exhibit several distinctive features. In this scenario supersymmetry
breaking at low energies is generally preferred. A firm prediction is the
existence of new states at or close to the TeV scale that are required by the
global symmetry. In the simplest unified models these have the quantum numbers of
the $10$ and $\overline{10}$ representations of SU(5). Another characteristic
feature is the existence of a Z' gauge boson with mass in the 300 GeV to a TeV
range, which has direct couplings to quarks and leptons. There are also hints as
to the superpartner spectrum. It is most natural for all the scalar superpartners
to be significantly heavier than the lightest neutral Higgs. Further the
electroweak gauginos are constrained to be lighter than the stops.  In contrast to
most other models which address the supersymmetric little hierarchy problem, the
lightest neutral Higgs is expected to be lighter than 130 GeV just as in the MSSM.

How well does this proposal address the supersymmetric little hierarchy problem?
This clearly depends on the details of the superpartner spectrum. Preliminary
numerical results in a gauge mediated model seem to indicate that the fine-tuning
is ameliorated by as much as an order of magnitude, from the 1-2 $\%$ level to
about the 10$\%$ level. Detailed investigation into this and other models is
currently underway {\cite{BCG}}.

\bigskip

\noindent {\bf Acknowledgements} \\ 
Z.C. would like to thank Ann Nelson for a very useful comment. A.B. 
would like to
thank Lawrence Berkeley National Laboratory and Cornell University for their kind
hospitality during completion of this work.  Z.C. and M.K.G. are supported by the
Director, Office of Science, Office of High Energy and Nuclear Physics, of the U.
S. Department of Energy under Contract DE-AC03-76SF00098, and by the National
Science Foundation under grant PHY-00-98840.

\end{document}